\documentclass[reprint, prb, citeautoscript, superscriptaddress]{revtex4-1}
\usepackage{amsmath}    % need for subequations
\usepackage{graphicx}   % need for figures
\usepackage{verbatim}   % useful for program listings
\usepackage{color}      % use if color is used in text
\usepackage{subfigure}  % use for side-by-side figures
\usepackage{hyperref}   % use for hypertext links, including those to external documents and URLs
\usepackage{gensymb}
\usepackage{amssymb}
\usepackage{mciteplus}
\usepackage[export]{adjustbox}
\begin{document}

\title[Nanowires]{Low-power photothermal self-oscillation of bimetallic nanowires}
\author{Roberto De Alba}
\author{T. S. Abhilash}
\affiliation{Department of Physics, Cornell University, Ithaca NY, USA}
\author{Richard H. Rand}
\affiliation{Department of Mathematics, Cornell University, Ithaca NY, USA}
\affiliation{Sibley School of Mechanical and Aerospace Engineering, Cornell University, Ithaca NY, USA}
\author{Harold G. Craighead}
\affiliation{School of Applied and Engineering Physics, Cornell University, Ithaca NY, USA}
\author{Jeevak M. Parpia}
\affiliation{Department of Physics, Cornell University, Ithaca NY, USA}
\email{jmp9@cornell.edu}

%\documentclass[journal=jacsat,manuscript=article, layout=twocolumn]{achemso}
%%\documentclass[11pt, oneside]{article} 
%
%\usepackage[version=3]{mhchem} % Formula subscripts using \ce{}
%\usepackage[T1]{fontenc}       % Use modern font encodings
%
%\usepackage{gensymb}
%\usepackage{amsmath}
%\usepackage{amssymb}
%
%\author{Roberto De Alba}
%\author{T. S. Abhilash}
%\affiliation[Deptartment of Physics]{Department of Physics, Cornell University, Ithaca NY, USA}
%\author{Richard Rand}
%\affiliation[Department of Mathematics]{Department of Mathematics, Cornell University, Ithaca NY, USA}
%\author{Harold G. Craighead}
%\affiliation[School of Applied and Engineering Physics]{School of Applied and Engineering Physics, Cornell University, Ithaca NY, USA}
%\author{Jeevak Parpia}
%\affiliation[Department of Physics]{Department of Physics, Cornell University, Ithaca NY, USA}
%\email{jmp9@cornell.edu}
%
%\title[Nanowires]
%	{Low-power photothermal self-oscillation of bimetallic nanowires}
%
%\abbreviations{IR,NMR,UV}
%\keywords{American Chemical Society, \LaTeX}
%
%\begin{document}

\begin{abstract}
%\singlespacing
We investigate the nonlinear mechanics of a bimetallic, optically absorbing SiN-Nb nanowire in the presence of incident laser light and a reflecting Si mirror. Situated in a standing wave of optical intensity and subject to photothermal forces, the nanowire undergoes self-induced oscillations at low incident light thresholds of $<1\, \rm{\mu W}$ due to engineered strong temperature-position ($T$-$z$) coupling. Along with inducing self-oscillation, laser light causes large changes to the mechanical resonant frequency $\omega_0$ and equilibrium position $z_0$ that cannot be neglected. We present experimental results and a theoretical model for the motion under laser illumination. In the model, we solve the governing nonlinear differential equations by perturbative means to show that self-oscillation amplitude is set by the competing effects of direct $T$-$z$ coupling and $2\omega_0$ parametric excitation due to $T$-$\omega_0$ coupling. We then study the linearized equations of motion to show that the optimal thermal time constant $\tau$ for photothermal feedback is $\tau \to \infty$ rather than the widely reported $\omega_0 \tau = 1$. Lastly, we demonstrate photothermal quality factor ($Q$) enhancement of driven motion as a means to counteract air damping. Understanding photothermal effects on micromechanical devices, as well as nonlinear aspects of optics-based motion detection, can enable new device applications as oscillators or other electronic elements with smaller device footprints and less stringent ambient vacuum requirements.
\end{abstract}

\maketitle

%\singlespacing
Micro- and nano-mechanical resonators are widely studied for applications including electro-mechanical circuit elements and sensing of ultra-weak forces~\cite{Gavartin2012b}, masses~\cite{Chaste2012}, and displacements~\cite{Teufel2011b}. An integral part of these systems is the detection method employed to readout motion, which must itself be extremely sensitive and %on some level
inevitably imparts its own force on the resonator, influencing the dynamics. The phase relation between mechanical motion and the resulting detector back-action determines whether this interaction will serve to dampen vibrations or amplify them, potentially leading to self-oscillation if the detector supplies enough energy per cycle to overcome mechanical damping.

Feedback due to external amplifiers has been used to generate self-oscillation of micro-mechanical resonators~\cite{Feng2008, Weldon2010, Villanueva2011, Chen2013a, Chen2016}; in such systems the oscillation amplitude $R$ is set either by nonlinearity of the amplifier or of the resonator. Systems in which mechanical motion influences the amount of laser light circulating in an optical cavity~\cite{Zalalutdinov2001, Aubin2004, Arcizet2006a, Metzger2008} or magnetic flux through a Superconducting QUantum Interference Device~\cite{Poot2010, Etaki2013} (SQUID) have also been shown to self-oscillate under the right experimental conditions. In these systems $R$ is set largely by the periodicity of the detection scheme -- either $R\approx {\lambda}/{4}$ where $\lambda$ is the laser wavelength or $R\approx {\Phi_0}/{2}$ where $\Phi_0$ is the displacement needed to change the SQUID flux by one flux quantum. In the case of a mechanical resonator coupled to an optical cavity, back-action can arise either from radiation pressure or photothermal force -- that is, thermally-induced deflection caused by optical absorption. The effects of these two forces are identical if the cavity resonance (with frequency $\Omega_\mathrm{c}$ and width $\kappa$) is sufficiently broad~\cite{Arcizet2006a, Kleckner2006, Gigan2006, Metzger2008, Eichenfield2009}; however if $\kappa$ is much smaller than the mechanical vibration frequency $\omega_\mathrm{m}$ the optomechanical system is said to be in the ``sideband-resolved regime,'' and radiation-pressure effects are enhanced at laser frequencies of $\Omega_\mathrm{c} \pm \omega_\mathrm{m}$.~\cite{Schliesser2009, Aspelmeyer2014} Radiation-pressure-based feedback with red detuning $(\Omega_\mathrm{c} - \omega_\mathrm{m})$ is currently one of the most promising experimental techniques for suppressing thermal motion and thereby accessing quantum behavior in mechanical systems.~\cite{Chan2011} Such low-$\kappa$ optical systems can, however, be difficult to attain and miniaturize.

\begin{figure*}[ht!]
   \centering
   \includegraphics[scale=0.5]{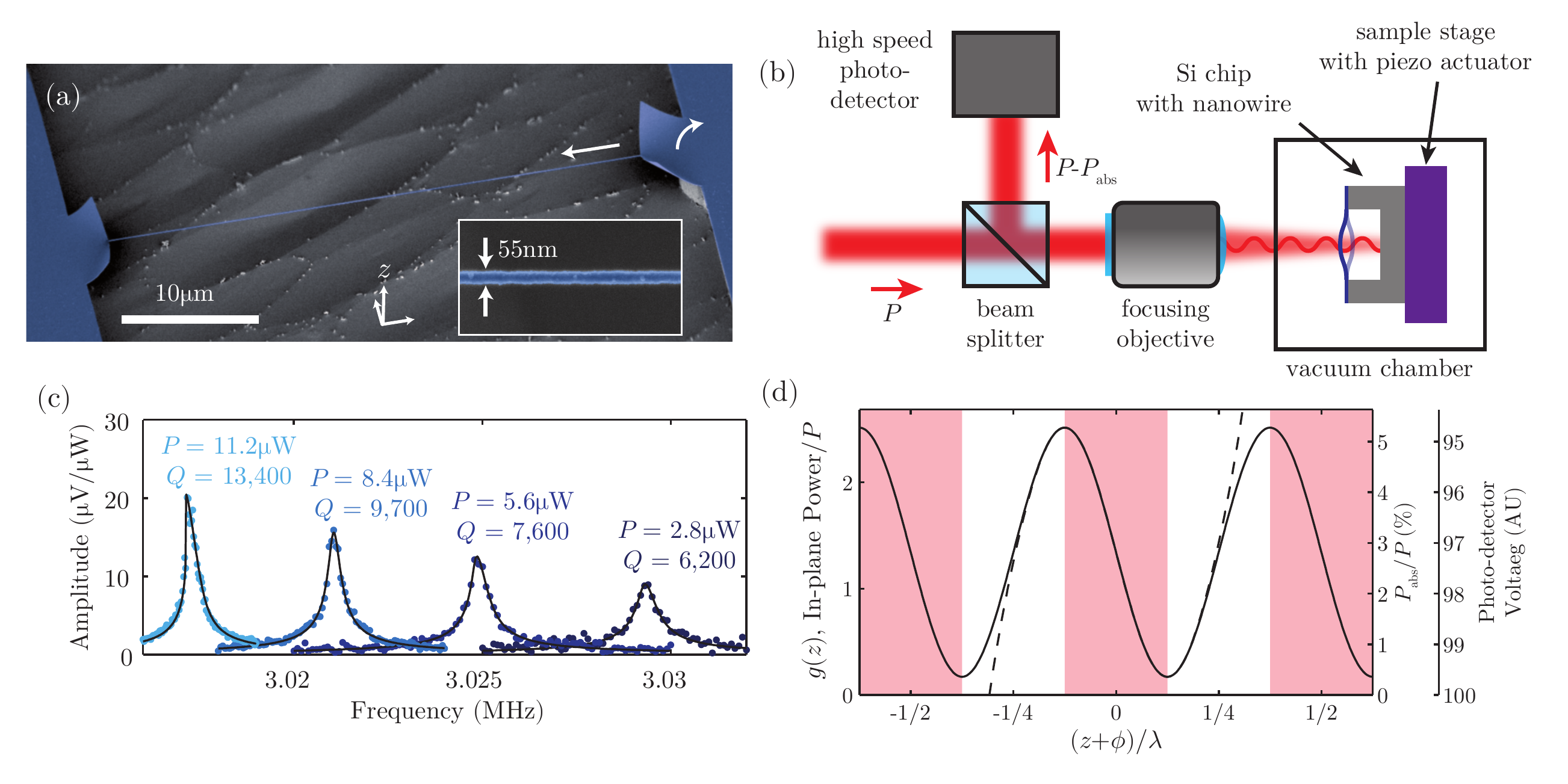} % requires the graphicx package
   \caption{The optomechanical system and experimental setup. (a) False-color scanning electron micrograph of our suspended device; blue: the SiN/Nb bilayer. Arrows indicate the competing tensile force and bimetallic ``torque'' that provide ${dz}/{dT}$ coupling. Inset: magnified top-down image of the nanowire. (b) The experimental setup: nanowire absorption modulates the reflected laser power, which is recorded by a high-speed photo-detector. (c) Nanowire resonance at laser powers below the threshold for self-oscillation, driven inertially by a piezo actuator; solid lines are Lorentzian fits. Considerable frequency softening ${d\omega_0}/{dT}$ and $Q$-enhancement can be seen as $P$ increases. (d) The optical intensity profile $g(z)$ versus distance $z+\phi$ to the Si mirror. Because the nanowire is much narrower than the incident laser beam, only $\approx 3\%$ of laser light interacts with the nanowire; of this $3\%$, the nanowire absorbs $\approx 70\%$. Self-oscillation occurs if the static nanowire is located in a shaded region and the power $P$ is sufficiently high. A dashed line indicates the Taylor-series approximation for $g(z)$ used in the perturbation theory.}
   \label{fig3.1}
\end{figure*}

Photothermal feedback places less stringent requirements on the optical system (as we show in this work), and has been explored in a broad range of mechanical device geometries through experiment \cite{Zalalutdinov2001, Aubin2004, Metzger2008a, Metzger2008, Ramos2008, Barton2012a, Ramos2012}, simulation \cite{Blocher2012, Blocher2013}, and theoretical studies \cite{Aubin2004, Metzger2008a, Restrepo2011}. While these works provide many insights into the underlying physics, some neglect the thermally-induced change in resonator equilibrium position $z_0$, while others neglect the change in resonant frequency $\omega_0$. In this work we have developed bimetallic nanowires that are designed to be especially susceptible to the photothermal force -- devices in which optically-induced changes to $z_0$ and $\omega_0$ cannot be neglected. Temperature-position coupling ${dz}/{dT}$ is provided by supporting bimetallic cantilevers at either end of the nanowire (shown in Fig.~\ref{fig3.1} (a)), and induces self-oscillation as well as changes in $z_0$. At room temperature these cantilevers apply an upward torque on the nanowire and change its $z$ position when its tension changes due to thermal expansion. Temperature-frequency coupling ${d\omega_0}/{dT}$, also due to thermal expansion, produces an overall shift in $\omega_0$ (Fig.~\ref{fig3.1} (c)) and modifies motion through $2\omega_0$ parametric excitation of the resonant frequency. We adapt the perturbation theory first discussed in Ref.~\citenum{Aubin2004}, and present our results for a general optical intensity profile $g(z)$. We then linearize the governing coupled $z,T$ equations to study nanowire behavior at laser powers below the threshold for self-oscillation.

Our optomechanical system is depicted in Figure \ref{fig3.1} (a,b). The nanowire has dimensions of $\sim(50\, \mathrm{nm})^2\times 40\, \mathrm{\mu m}$ and is suspended $8\, \rm{\mu m}$ above a Si back-plane. Incident laser light (beam diameter $d_L \approx 2.5\, \rm{\mu m}$) is focused near the wire center, and reflects off of the underlying Si to form a standing wave of optical intensity; our one-mirror optical system thus functions similarly to a very low-finesse two-mirror cavity. The total optical power (or more precisely, the electric field energy density $|\vec{E}(z)|^2$) in a plane parallel to the mirror at a distance $z$ is given by $P g(z)$, where $g(z)$ is the dimensionless intensity profile and $P$ is the incident beam power; all $P$ values given throughout this work signify this total beam power. Because the nanowire is extremely narrow, it covers only $\approx 3\%$ of the incident beam by area and is therefore assumed not to influence $g(z)$. It does, however, absorb a small portion of the local power $P_\mathrm{abs}$, and nanowire motion generates fluctuations in the reflected laser beam that can be measured using a high-speed photodetector. The detected signal is proportional to $P-P_\mathrm{abs}$, as shown in Fig.~\ref{fig3.1} (d). This detection method has the benefit of utilizing the same light which induces self-oscillation, but is highly nonlinear for oscillation amplitudes $R\gtrsim {\lambda}/{8}$, where $\lambda=660\, \rm{nm}$ is the laser wavelength used. If the optical field profile $g(z)$ is known, this detector nonlinearity can be used to deduce the absolute size of mechanical motion.

\begin{figure}[h!]
   \centering
   \includegraphics[scale=0.4, center]{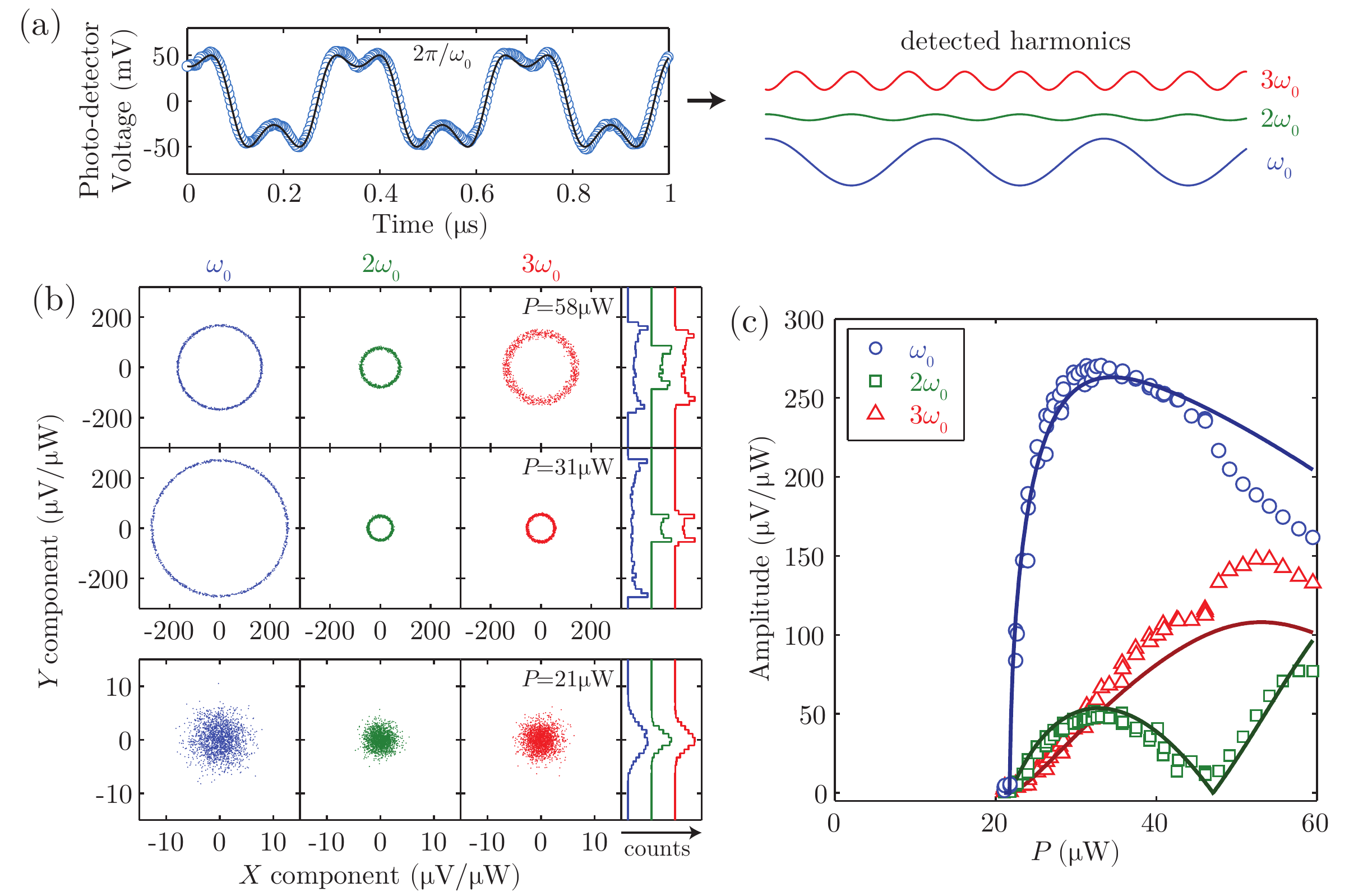} % requires the graphicx package
   \caption{Photothermal self-oscillation. (a) Measured photo-detector signal during nanowire self-oscillation (circles), and its decomposition into Fourier components (solid lines). Although the nanowire motion is a near-pure sinusoid, the nonlinear optical readout results in detected harmonics at integer multiples of the oscillation frequency. (b) Phase portraits of undriven nanowire motion as measured in the frequency domain by a multi-channel lock-in amplifier centered about the resonant frequency. $X$ and $Y$ denote cosine and sine components of motion. The critical power needed for self-oscillation is $P_\mathrm{crit}=22\, \rm{\mu W}$. Data below this power (lowest row) is a combination of thermal motion and detector noise, while data above this power (upper rows) has a well-defined nonzero amplitude. (c) Data points: amplitudes of the self-oscillation signals shown in (b) versus laser power $P$. Solid lines are a best fit based on the IPT model described in the text.}
   \label{fig3.2}
\end{figure}

Self-oscillation of the nanowire is shown in Figure \ref{fig3.2}. Measured in the time domain (Fig.~\ref{fig3.2} (a)), purely sinusoidal motion with $R\approx {\lambda}/{4}$ results in a detected signal that saturates as $z$ traverses the extremes of $g(z)$. This results in detected harmonics of the vibration frequency that can readily be measured in the frequency domain. Figure \ref{fig3.2} (b) shows the nanowire motion as measured by a multi-channel lock-in amplifier whose reference frequency is centered at the resonant frequency $\omega_0 \approx 2\pi \times 3\, \rm{MHz}$ with a $10\, \rm{kHz}$ bandwidth for three different laser powers; the three harmonics shown ($\omega_0, 2\omega_0, 3\omega_0$) were measured simultaneously. The reference frequency was adjusted at each power to follow the resonance. Nanowire motion is plotted as $\rm{X}$ and $\rm{Y}$ quadratures, or in-phase and out-of-phase components relative to a fixed phase. The lower panel displays nanowire motion just below the critical power ($P_\mathrm{crit} =22\, \rm{\mu W}$), which is a combination of thermal motion and electrical noise about the origin; this has the expected Gaussian distribution. As $P$ is increased above $P_{\rm{crit}}$, all three harmonics demonstrate sharply-defined nonzero amplitudes. This optically-induced motion has a phase that randomly cycles through all possible angles as time progresses at nearly constant amplitude. All plots show 1,000 data points except for the lower panels which each contain 2,000 points. Figure \ref{fig3.2} (c) shows the amplitude of these harmonics ($\sqrt{X^2+Y^2}$) for many values of $P$. Solid lines are a best fit (with a total of 4 free parameters) based on the model described below. Deviation of the fit at high powers could be due to aberrations of the optical plane wave $g(z)$ caused by the nanowire, as studied previously by Refs.~\citenum{Ramos2012}~\&~\citenum{Ramos2013}. All measured signals are normalized by $P$, the laser power used.

The governing differential equations for the position and temperature of our photothermal system~\cite{Aubin2004} are:
\begin{align}
	\label{eq3.1}
	& \ddot{z} + \gamma\dot{z} + \omega_0^2 (1+C T)(z - DT) = 0 \\
	\label{eq3.2}
	& \dot{T} + \frac{1}{\tau} T = PA g(z)
\end{align}
Here $\omega_0, \gamma$ are the intrinsic resonant frequency and damping of the nanowire. $T$ denotes the temperature above ambient and $C, D$ are the changes in resonator frequency and position per unit temperature, respectively. The second equation is Newton's law of cooling, where $\tau$ denotes the thermal diffusion time constant, and the right-hand-side describes heat absorption from the incident laser. $A$ includes the thermal mass and optical absorption of the nanowire, as well as its $\approx 3\%$ area coverage of the incident laser beam. Detailed calculations of the thermal parameters in Eqs.~\ref{eq3.1} \&~\ref{eq3.2} based on the  materials and dimensions of our system are presented in the Supplementary Information.

Because the nanowire does not interact appreciably with the incident laser, we can approximate the optical field to be:
\begin{equation}
	g(z) = \alpha + \beta \sin^2 \left( \frac{2\pi (z+\phi)}{\lambda} - \frac{\pi}{4} \right)
	\label{eq3g}
\end{equation}
Here $\alpha, \beta$ are determined by the reflection coefficient of the Si back-plane, and $\phi$ is the $P=0$ nanowire position within the standing wave. The factor of $-\pi/4$ is added to center the self-oscillation region (negative ${dg}/{dz}$ region, Fig.~\ref{fig3.1} (d)) about $z+\phi=0$. The total mirror-nanowire distance is $z+\phi+(\lambda/2)(n-1/4)$, where the integer $n$ is irrelevant to our measurements. % n\approx 24

In other device geometries, large mechanical resonators can generate significant internal and external optical reflections, producing a Fabry-Perot interference effect which results in $g(z)$ having sharper peaks and wider valleys, or skewing its peaks left or right. For this reason we present our theoretical results for a general intensity profile $g(z)$. In all cases, however, $g(z)$ is periodic in ${\lambda}/{2}$.

During self-oscillation, the resonator position is well modeled by $z(t)=z_0 + R\cos(\omega t)$ where $z_0$ is the temperature-dependent equilibrium position. This value can be estimated by solving Eqs.~\ref{eq3.1} \&~\ref{eq3.2} for the case of a static nanowire, which give the implicit equation ${z_0} = \tau D PA g(z_0)$. Near $P=0$ this formula has only one solution for $z_0$, but more solutions become available as $P$ increases. For high enough $P$ values, solutions nearest $z=0$ can cease to be valid; this suggests that the static wire exhibits discontinuous jumps in $z_0$ as $P$ is increased quasi-statically. The static solution to Eqs.~\ref{eq3.1} \&~\ref{eq3.2} is studied further in the Supplementary Information. While the static solution for $z_0$ (and the corresponding temperature $T_0 = {z_0}/{D}$) is a useful starting point for analyzing the self-oscillating nanowire, in what follows we will show that typical oscillation amplitudes $R$ produce sizable changes in $T_0$ (and $z_0$).

Although Eqs. \ref{eq3.1}-\ref{eq3g} are nonlinear and cannot be solved exactly, perturbative methods can be applied. Here we employ the Poincar\'e-Lindstedt method, which requires scaling $\gamma, C,$ and $D$ in Eq.~\ref{eq3.1} by a small dimensionless parameter $\varepsilon \ll 1$. Eqs.~\ref{eq3.1} \&~\ref{eq3.2} can then be solved for $z(t),T(t)$, and $\omega_1$ (the self-oscillation frequency) to any desired order in $\varepsilon$. The method also requires approximating $g(z)$ by the first few terms of its Taylor series. We expand $g(z)$ about $z+\phi=0$ and keep enough terms such that the optical field is accurately modeled over an entire period $|z+\phi|< \lambda/4$:
\begin{equation}
\begin{aligned}
	g(z) \approx k_0 + k_1(z+\phi) + k_3(z+\phi)^3 \\ + k_5(z+\phi)^5 + k_7(z+\phi)^7
%	g(z) \approx \left(\alpha + \frac{\beta}{2} \right) -2\pi \beta(z+\phi) +\frac{16}{3}\pi^3 \beta(z+\phi)^3 \\
%		-\frac{64}{15}\pi^5 \beta (z+\phi)^5 + \frac{512}{315} \pi^7 \beta (z+\phi)^7
	\label{eq3g2}
\end{aligned}
\end{equation}
where $k_0=(\alpha+\beta/2)$, $k_1 =-2\pi\beta$, $k_3=(16/3)\pi^3\beta$, $k_5=-(64/15)\pi^5\beta$, and $k_7=(512/315)\pi^7\beta$. A comparison of this approximation with the exact $g(z)$ is shown in Fig.~\ref{fig3.1} (d). The perturbation theory is presented in its entirety in the Supplementary Information, but the main results are given below.

Using Eq. \ref{eq3g2} and solving Eqs.~\ref{eq3.1} \&~\ref{eq3.2} to order $\varepsilon^1$ gives the following equation for $R$:
\begin{equation}
	%c_1R^6 + c_2R^4 + c_3 R^2 + c_4 = 0
	0 = c_0 + c_1 R^2 + c_2 R^4 + c_3 R^6
	\label{eq3R}
\end{equation}
where
\begin{align*}
%	\label{eq3c1}
	c_0 & = \frac{\omega_1^2 D}{1 + \omega_1^2 \tau^2}  \, g^{(1)}_{z_0} 
		+ \frac{\gamma}{\tau^2 PA} \\
%	c_{n\geq 1} & = \frac{1}{2^{2n}(n+1)!} \Bigg[
%	 \frac{\omega_1^2 D}{1 + \omega_1^2 \tau^2} \, \frac{ g^{(2n+1)}_{z_0}}{n!} \\
%		& \hspace{75pt} -\frac{\omega_0^2 C}{1 + 4\omega_1^2 \tau^2} \, 
%		\frac{2 g^{(2n)}_{z_0}}{(n-1)!} \Bigg] \\
	c_1 & = \frac{\omega_1^2 D}{1 + \omega_1^2 \tau^2}  \, \frac{ g^{(3)}_{z_0}}{2^2 1! 2!} 
		- \frac{\omega_0^2 C}{1 + 4\omega_1^2 \tau^2} \, \frac{g^{(2)}_{z_0}}{2^1 0! 2!} \\
	c_2 & = \frac{\omega_1^2 D}{1 + \omega_1^2 \tau^2}  \, \frac{ g^{(5)}_{z_0}}{2^4 2! 3!} 
		- \frac{\omega_0^2 C}{1 + 4\omega_1^2 \tau^2} \, \frac{g^{(4)}_{z_0}}{2^3 1! 3!} \\
	c_3 & = \frac{\omega_1^2 D}{1 + \omega_1^2 \tau^2}  \, \frac{ g^{(7)}_{z_0}}{2^6 3! 4!} 
		- \frac{\omega_0^2 C}{1 + 4\omega_1^2 \tau^2} \, \frac{g^{(6)}_{z_0}}{2^5 2! 4!}
	\label{eq3c4}
\end{align*}
Here we have introduced $\omega_1^2 = \omega_0^2(1 + CT_0)$ as the new resonant frequency and $g^{(n)}_{z_0}$ as the $n^{\rm{th}}$ derivative of $g(z)$ evaluated at $z = z_0$. This result is hereafter referred to as the First Order Perturbation Theory (FOPT) solution. The number of terms in Eq. \ref{eq3R} increases if more terms are kept in the Taylor expansion Eq. \ref{eq3g2} (following the clear pattern in $c_0\dots c_3$), however the terms shown are sufficient to accurately model our experimental data.

Eq.~\ref{eq3R} indicates that $R$ is influenced by both the temperature-position coupling $D$ and the temperature-frequency coupling $C$. Interestingly, $D$ influences self-oscillation via temperature fluctuations at the oscillation frequency $\omega_1$, while $C$ does so via temperature fluctuations at $2\omega_1$; the effect of $C$ is thus equivalent to parametric $2\omega_1$ excitation of the resonant frequency. Eq.~\ref{eq3R} also suggests that as $z_0$ changes, $C$ dominates near points of $g\left(z_0\right)$ with even symmetry (extrema) while $D$ dominates near points with odd symmetry (inflection points). The threshold for self-oscillation occurs when $R=0$ in Eq.~\ref{eq3R}; this leads to $c_0=0$ and gives a critical laser power of:
\begin{equation}
	P_\mathrm{crit} = -\frac{\gamma \left( 1 +  \omega_1^2 \tau^2 \right)} {\omega_1^2 \tau^2 DA g^{(1)}_{z_0} }
	\label{eq3Pcrit}
\end{equation}
This expression reveals the source of low critical power in our nanowire: a combination of low thermal mass $A$, long thermal time constant $\omega_1\tau \approx 400$, and large coupling $D=1.64\, \rm{nm/\degree C}$ afforded by our cantilevers. Further, because $\gamma, D, A$ are all positive, a negative optical gradient is needed for self-oscillation. While the sensitivity of $P_\mathrm{crit}$ on $\tau$ is rather weak for $\omega_1 \tau >1$, it is noteworthy that short time constants $\tau \to 0$ inhibit self-oscillation. We revisit this later in the paper where we discuss operation of the wires in the presence of $\mathrm{N_2}$ gas. For the case $D=0, C\neq0$ Eq.~\ref{eq3R} still supports limit cycle oscillations, but has no $R=0$ solution. This suggests that $z(t) = z_0$ remains a stable equilibrium point even for $P>P_\mathrm{crit}$, and only initial conditions of $(z,\dot{z})$ sufficiently close to $z(t) = z_0 + R\cos(\omega t)$ will lead to oscillation. One can therefore draw an attractor diagram to describe which initial conditions lead to limit cycle behavior and which approach the stable equilibrium \cite{Metzger2008}.

As mentioned above, FOPT predicts a change in the time-averaged temperature of the nanowire due to self-oscillation. This addition to $T_0$ is
\begin{equation}
	\delta T_0 = -T_0 + \tau PA \sum_{n=0}^3 \frac{R^{2n} g^{(2n)}_{z_0}} {2^{2n} (n!)^2}
	\label{eq3dT0}
\end{equation}
The nanowire equilibrium position thus relocates to $z_0 = D\left( T_0 + \delta T_0 \right)$ during self-oscillation. 
Although one could proceed to order $\varepsilon^2$ in perturbation theory to account for this equilibrium shift, the resulting algebraic expressions quickly become cumbersome. An approach that is easier to implement and was used to fit the data in Fig.~\ref{fig3.2} (c) is to recursively perform FOPT while updating $T_0$ and $ z_0$ with successive $\delta T_0$ values. Starting with the static nanowire solution ($z_0 = \tau D P A g(z_0)$), $R$ and $\delta T_0$ are iteratively calculated until $R$ converges on a fixed value and $\delta T_0$ converges on zero. 
This scheme is hereafter referred to as Iterated Perturbation Theory (IPT). We find in practice that IPT converges most reliably if $\delta T_0$ is multiplied by a small scaling factor ($0.05$ was used) before being added to $T_0$; convergence typically occurs within 20-100 iterations.

\begin{figure}[t!]
   \centering
   \includegraphics[scale=0.4, center]{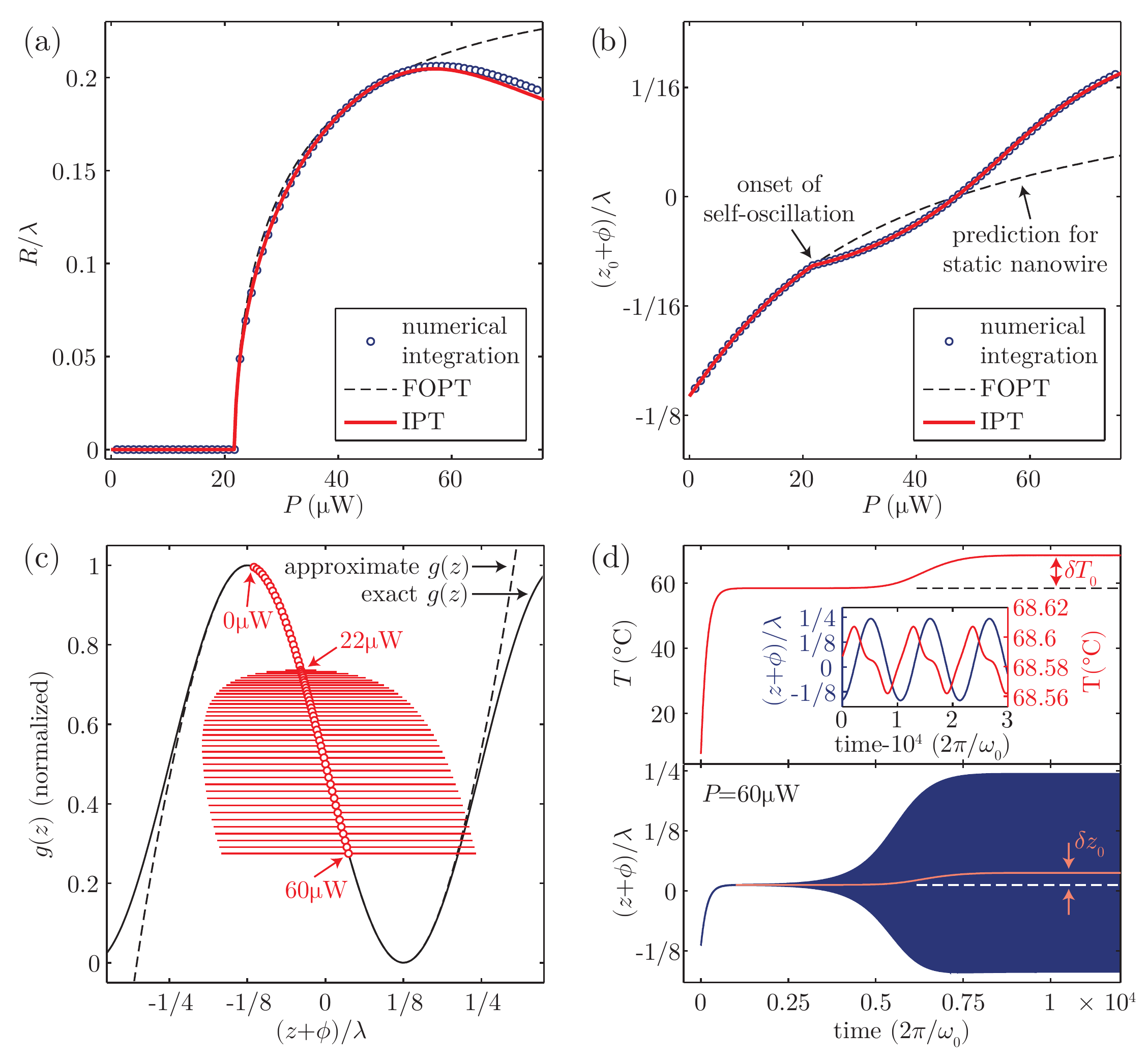} % requires the graphicx package
   \caption{Detailed behavior of the nanowire according to fits of the experimental data. (a,b) Comparison of the oscillation amplitude $R$ and equilibrium position $z_0$ calculated by perturbation theory and numerical integration, with $\phi/\lambda = -0.114$. Note that $z_0=0$ at $P=0$. The shift in $z_0$ due to self-oscillation is clearly visible in (b). (c) Nanowire position within the optical field $g(z)$ as $P$ increases. Red points (spaced every $1\, \rm{\mu W}$) indicate the changing $z_0$ value, while horizontal lines indicate the extent of $R$. (d) Numerical integration results at $P=60\, \rm{\mu W}$ with the initial condition $(z,\dot{z},T) = (0,0,0)$; only the upper and lower envelopes of oscillation are shown. In the lower panel, a solid line signifies the peak-peak moving average, which is an indication of $z_0$. The shift in $z_0$ after $t=5\times10^3$ closely follows the trend in $T(t)$ shown in the upper panel. Inset: magnified image of these results near $t = 10^4$, showing the harmonic content of $z(t)$ and $T(t)$.}
   \label{fig3.3}
\end{figure}

A comparison of FOPT, IPT, and numerical integration of Eqs. \ref{eq3.1}--\ref{eq3g} is shown in Figure~\ref{fig3.3} (a,b). The parameters used are derived from the IPT fit to experimental data in Fig.~\ref{fig3.2} (c) -- in this fit the only free parameters were $\phi, \tau,d_L$, and an overall vertical scaling factor. It should be noted that while IPT reproduces the results of numerical integration almost exactly, the former required only $\sim$1 second of computation time while the latter required $4-5$ hours. Fig.~\ref{fig3.3} (c) shows the nanowire position as it moves through the optical field. The deviation of $z_0$ away from its static value due to $\delta T_0$ is clearly visible in Fig.~\ref{fig3.3} (b). Interestingly, $z_0$ trajectories from numerical integration and static theory intersect at $z_0 + \phi = 0$ -- \textit{i.e.} at the inflection point of $g(z)$; here the odd symmetry of $g(z)$ results in $\delta T_0 = 0$ in Eq.~\ref{eq3dT0}. The inflection point is crossed by $z_0$ at $P\approx 47\, \rm{\mu W}$, while the maximum $R$ value occurs at the slightly higher power of $P\approx 56\, \rm{\mu W}$.

As shown in the numerical integration results of Fig.~\ref{fig3.3} (d), self-oscillation requires roughly $10^4$ oscillation cycles to reach steady state at $P=60\, \rm{\mu W}$. We note that this ``equilibration time'' drastically increases for $P$ values approaching $P_\mathrm{crit} = 22\, \rm{\mu W}$; a maximum of $3\times 10^5$ cycles were required just above the transition. Also shown in Fig.~\ref{fig3.3} (d), the shift $\delta z_0 = 0.0251 \lambda$ due to self-oscillation exactly matches the observed change in temperature $\delta T_0 = \delta z_0 / D = 10.1\degree\rm{C}$, where $D= 1.64\, \rm{nm / \degree C}$ for this system. The numerical results in the Fig.~\ref{fig3.3} (d) inset show that during self oscillation $z(t)$ is a nearly pure tone at frequency $\omega_1$. A Fourier series fit to this data (not shown) reveals that the next largest harmonic component is $2\omega_1$, with  $0.001\%$ the amplitude of $\omega_1$ motion. It is the pureness of this tone that leads to the excellent agreement between numerical integration and IPT -- after all the perturbation theory is predicated on the assumption $z(t) = z_0 + R\cos(\omega_1 t)$. Numerical results for $P> 60\, \rm{\mu W}$ reveal that higher harmonics of the $\omega_1$ motion grow steadily as $P$ increases ($2\omega_1$ reaching $0.004\%$ at $80\, \rm{\mu W}$), possibly explaining the growing deviation from IPT seen in Fig.~\ref{fig3.3} (a). The oscillation frequency in the Fig.~\ref{fig3.3} (d) inset is $0.93\, \omega_0$, in close agreement with the expected $\omega_1 = \omega_0 \sqrt{1 + C T_0}\approx 0.92\, \omega_0$, where $C=-0.0022$ and $T_0 = 68.58\degree\rm{C}$. The $1\%$ increase in frequency is likely due to $\omega_2$, the $\epsilon^1$-order correction to the oscillation frequency, which is calculated in the Supplementary Information.%Motion is expected to induce a slight frequency shift, as discussed  stiffening is predicted by the perturbation theory as presented in the Supplementary Information.

Perturbation theory can also be used to predict whether the onset of self-oscillation will be exhibit hysteresis. Such behavior is referred to as a subcritical Hopf bifurcation, and would manifest as a continuation of stable self-oscillation for some range of powers as $P$ is decreased below $P_\mathrm{crit}$. The distinction between a hysteretic or non-hysteretic transition (subcritical or supercritical bifurcation) depends upon whether $c_1$ in Eq.~\ref{eq3R} is negative or positive. Therefore
\begin{equation}
	\frac{\omega_1^2 D g^{(3)}_{z_0}} {1 + \omega_1^2 \tau^2} < \frac{2\omega_0^2 C g^{(2)}_{z_0}}{1 + 4\omega_1^2 \tau^2}
	\label{eq3hyst}
\end{equation}
is the necessary condition for hysteresis. Because $C<0$ in this experiment, we would expect hysteretic behavior when $z_0$ is near a maximum of $g(z)$. The width of the hysteresis region (\textit{i.e.} how low $P$ can be while still maintaining self-oscillation) is calculated in the Supplementary Information.

Lastly, we focus on the behavior of our nanowire for laser powers $P<P_\mathrm{crit}$. Since the vibration amplitude in this case is typically much smaller than $\lambda/4$, it suffices to approximate $g(z)$ by a linear expansion about $z=z_0$ in Eq.~\ref{eq3.2}: $g(z) \approx g(z_0) + g^{(1)}_{z_0}(z-z_0)$. Furthermore, we can neglect any time-dependent $CT$ terms in Eq.~\ref{eq3.1}. This then leads to the linearized equations
\begin{align}
	\label{eq14}
	& \ddot{x} + \gamma \dot{x} + \omega_1^2(x - Du) = f_d e^{i\omega t} \\
	& \dot{u} + \frac{1}{\tau} u = PA g^{(1)}_{z_0} x
	\label{eq15}
\end{align}
where we have introduced the new variables $x = z - z_0$, $u = T - T_0$ and added the driving term $f_d$ at frequency $\omega$. In this linearized system we can safely use the complex solutions $x = \tilde{x} e^{i\omega t}$ and $u = \tilde{u} e^{i\omega t}$. Based on Eq.~\ref{eq15}, these are related by $u = x\, ({\tau PA g^{(1)}_{z_0}})/({1 + i\omega \tau})$. Substituting this into Eq.~\ref{eq14} and collecting real and imaginary terms, one can recast the mechanical system as $\ddot{x} + \gamma_\mathrm{eff} \dot{x} + \omega_\mathrm{eff}^2 x = f_d e^{i\omega t}$ where the effective resonant frequency $\omega_\mathrm{eff}$ and damping $\gamma_\mathrm{eff}$ are:
\begin{align}
	\label{eq3weff}
	\omega_{\mathrm{eff}}^2 & = \omega_1^2 \left( 1 -\frac{ \tau DPA  g^{(1)}_{z_0}}{1 +\omega^2 \tau^2} \right) \\
	\gamma_\mathrm{eff} & = \gamma + \frac{\omega_1^2 \tau^2 DPA g^{(1)}_{z_0}} {1 + \omega^2 \tau^2}
	\label{eq3geff}
\end{align}
Firstly, we note that the photothermal terms in $\omega_\mathrm{eff}$ constitute a roughly 1 part in $10^6$ correction for the experimental parameters used in this work; thus to very good approximation $\omega_\mathrm{eff} = \omega_1$. Next, we should expect self-oscillation to occur when $\gamma_\mathrm{eff}=0$. Substituting $P=P_\mathrm{crit}$ from perturbation theory (Eq.~\ref{eq3Pcrit}) and $\omega=\omega_1$ indeed gives $\gamma_\mathrm{eff}=0$, showing compatibility of these two models.
Interestingly, the photothermal damping shift on resonance is $\Delta \gamma = |\gamma_\mathrm{eff} -\gamma| \propto \frac{\tau^2}{1+\omega_1^2 \tau^2}$, which increases monotonically as $\tau\to\infty$. Long time constants $\omega_1\tau \gg 1$ therefore strengthen the photothermal effect. This can also be seen by setting $\frac{1}{\tau} = 0$ in Eq.~\ref{eq15}, which 
results in $u\propto ix$. In this case, $u$ is perfectly out of phase with $x$, meaning it contributes entirely to damping in Eq.~\ref{eq14}.

These results appear to be counter to those of previous theoretical studies which model the photothermal effect as a time-delayed back-action force $F(x)$ that responds to changes in $x$ after a time constant $\tau$.~\cite{Metzger2008a, Metzger2008} Such a model produces the result $\Delta \gamma \propto \frac{\tau}{1 + \omega^2 \tau^2} \frac{dF}{dx}$, which is maximized (in magnitude) when $\omega \tau = 1$ and vanishes as $\tau \to \infty$. The discrepancy here lies in ${dF}/{dx}$. Adapting our Eqs.~\ref{eq14} \&~\ref{eq15} to such a model reveals that the thermal force magnitude (\textit{i.e.} the asymptotic value after a change in $x$) is $F(x) = k Du = k D\tau PA g^{(1)}_{z_0} x$, where $k$ is the mechanical spring constant. This then leads to $\Delta \gamma \propto \frac{\tau^2}{1 + \omega^2\tau^2}$, in agreement with our earlier result.

\begin{figure}[t!]
   \centering
   \includegraphics[scale=0.4, center]{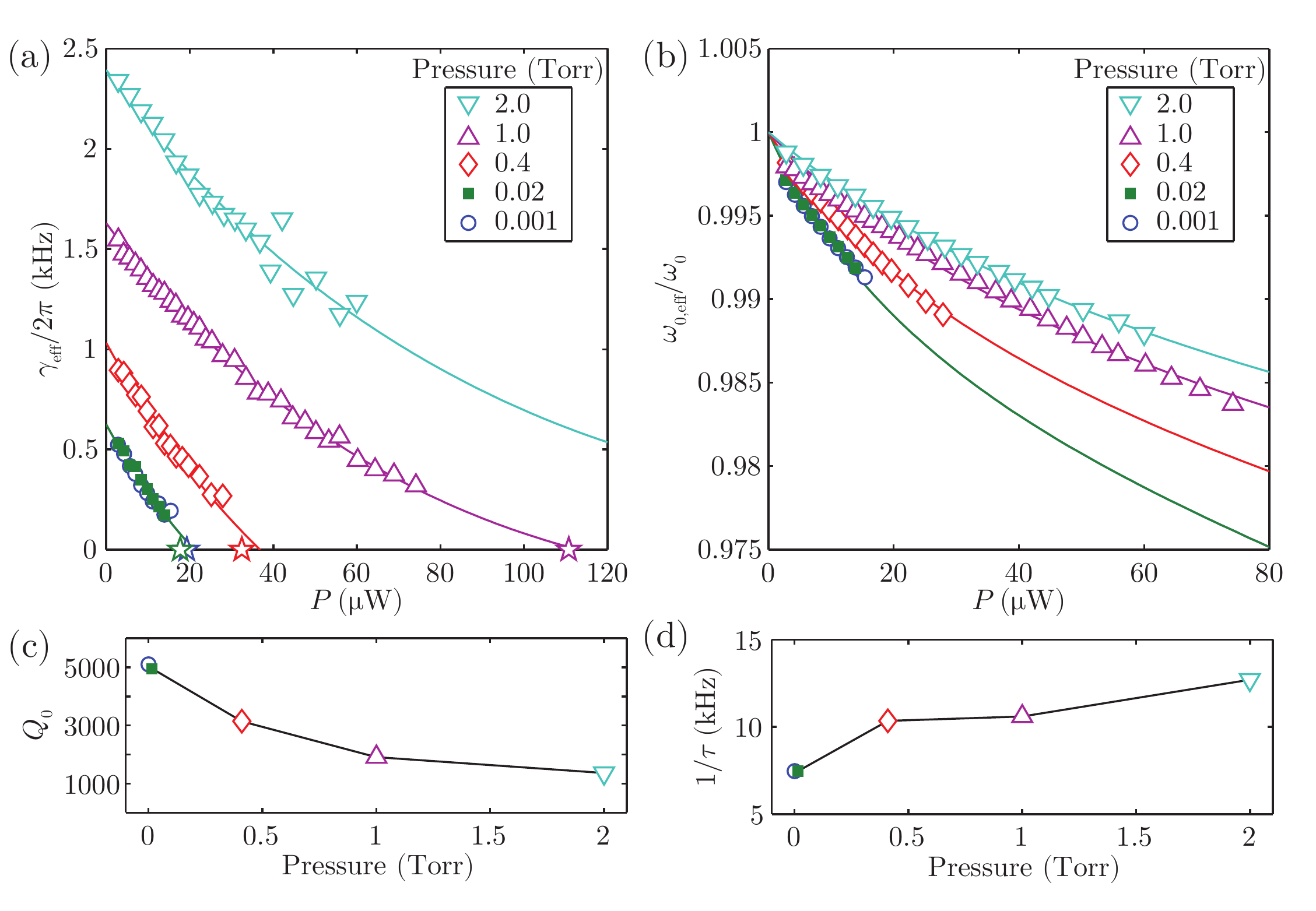} % requires the graphicx package
   \caption{Nanowire behavior for $P<P_\mathrm{crit}$ under various $\mathrm{N_2}$ pressures. (a,b) Nanowire effective damping $\gamma_\mathrm{eff}$ and resonant frequency $\omega_\mathrm{eff}$. These values were obtained from Lorentzian fits to piezo-driven resonance peaks such as those shown in Fig.~\ref{fig3.1} (c). Stars in (a) indicate the measured onset of self-oscillation. Solid lines are fits to Eqs.~\ref{eq3weff} \&~\ref{eq3geff}. (c) $Q$ factors at $P=0$ extrapolated from fits in (a,b). (d) Thermal diffusion rate $1/\tau$ versus gas pressure.}
   \label{fig3.4}
\end{figure}

We have experimentally tested Eqs.~\ref{eq3weff} \&~\ref{eq3geff} for several values of $\gamma$ as shown in Figure~\ref{fig3.4}. In these measurements $\gamma$ was varied by introducing pure $\rm{N_2}$ gas into our sample test chamber; doing so added drag to the nanowire motion, resulting in higher intrinsic damping $\gamma$ and lowered $Q$ factors (Fig.~\ref{fig3.4} (c)). All preceding measurements were performed with pressure $\ll 10^{-3}$ Torr. The fits shown in Fig.~\ref{fig3.4} (a,b) were constrained at the lowest two pressures to maintain consistent thermal parameters with the fit in Fig.~\ref{fig3.2} (c). At higher pressures $\tau$ was allowed to vary, as nanowire interaction with ambient gas likely increases its thermal dissipation rate. The laser waist diameter $d_L$ and initial optical field position $\phi$ were also allowed to differ from Fig.~\ref{fig3.2} (c) as each change in pressure required manual refocusing, and the roughness of the Si back-plane led to changes in $\phi$ based on exact laser positioning. Here $\phi/\lambda = 0.044$ compared to the value of $-0.114$ in Fig.~\ref{fig3.2} (c); $d_L = 2.5\, \rm{\mu m}$ for the two highest pressures and $d_L = 2.0\, \rm{\mu m}$ for all lower pressures.

Curvature in the $\gamma_\mathrm{eff}$ and $\omega_\mathrm{eff}$ fits is due to the changing equilibrium position $z_0$ as $P$ increases, and the resulting change in $g^{(1)}_{z_0}$. Because of this curvature, the $\gamma_\mathrm{eff}$ trajectory for $2.0\, \rm{Torr}$ is not expected to enter self-oscillation at higher $P$ values.  It is however possible that if $z_0$ can extend to the next negative region of $g^{(1)}_{z_0}$, near $z+\phi = \lambda/2$, $P$ would be large enough to support self-oscillation. We note that for the four values of pressure where self-oscillation is seen, the two lowest pressures yield $P_\mathrm{crit}\approx 22\, \rm{\mu W}$ identical to the value with no $\mathrm{N_2}$ gas added, and are consistent with the $Q$ value seen with no added gas. For the case of the two higher pressures, $0.4\, \mathrm{Torr}$ and $1\, \mathrm{Torr}$, the introduction of gas increases the damping (higher $\gamma_\mathrm{eff}$) and shortens the $\tau$, requiring an additional power to overcome damping. % At the highest pressure ($2\, \mathrm{Torr}$), the critical power is not achieved in our setup.
Above this pressure, self-oscillation cannot be reached in our present setup. Even so, the results of Fig.~\ref{fig3.4} demonstrate the capability of photothermal feedback to counteract air damping at low pressures. Such optical $Q$-enhancement could lower the stringent vacuum requirements of typical micro-electro-mechanical device applications.

We have presented an experimental and theoretical study of photo-thermal feedback in mechanical nanowires. While the device tested self-oscillates under the illumination of a $22\, \rm{\mu W}$ laser beam, only $\sim3\%$ of this beam is incident on the ultra-fine nanowire -- suggesting that incident powers of $<1\, \rm{\mu W}$ are ultimately necessary to induce motion. This is significantly lower than the $300\, \rm{\mu W}$ to few mW required in previously studied free-space photothermal structures~\cite{Aubin2004, Zalalutdinov2003, Ramos2012}, and lower still than the $\approx10\, \rm{\mu W}$ reported for an optical-cavity-coupled photothermal structure~\cite{Metzger2008}, where the two-mirror cavity results in much higher optical field gradients $dg/dz$. The low power needed in our system is attributable to the low thermal mass of the nanowire and large temperature-position coupling $D$ afforded by the supporting cantilevers. A simple beam-theory calculation suggests that $D$ scales with cantilever length $L$ and width $w$ as $L^3/w$ (see Supplementary Information), suggesting that even stronger photothermal effects can readily be achieved. We have observed that the equilibrium position $z_0$ of this system is strongly tunable with incident laser power and can drastically affect nanowire dynamics. Self-oscillation in this system is due in part to temperature oscillations at the vibration frequency $\omega_1$ and to parametric $2\omega_1$ oscillations of the resonant frequency. %We have also studied this system at low laser powers (below the self-oscillation threshold), demonstrating $Q$-enhancement in the face of low-pressure gaseous drag.
The perturbation theory used here can readily be adapted for systems in which micro-mechanical resonators are coupled to magnetic SQUID circuits, optical cavities, or other periodic external systems. 

It is well established that a self-oscillating system can become entrained if a sufficiently strong driving force is applied -- \textit{i.e.} the system will oscillate at the driver frequency rather than its own natural frequency.~\cite{Zalalutdinov2001, Pandey2006, Pandey2007, Blocher2013a} Such a system is promising for a number of electro-mechanical applications, including narrow bandpass filters and related electrical signal processing devices. Although we have observed such behavior in our nanowires (not shown here), further work is needed to extend the perturbation theory to predict the entrainment bandwidth as a function of driver strength and laser power.

\begin{acknowledgments}
We thank A. T. Zehnder for guidance in applying beam-theory to our supporting cantilevers, as well as other helpful discussions. This work was supported in part by the Cornell Center for Materials Research with funding from the NSF under DMR-1120296 and by the NSF under DMR-1202991.
\end{acknowledgments}

\bibliography{nwire_bib_4.bib}

\end{document}